\documentclass[Singlecolumn,aps,pre,showpacs,preprintnumbers]{revtex4}
\RequirePackage{ifpdf} 
\usepackage[dvipdf]{graphicx}
 \usepackage{xcolor, graphicx}
\usepackage{epstopdf}
\usepackage{epsfig} 
\usepackage{hyperref}
\usepackage{graphics}
\usepackage{color}      
\usepackage{dcolumn}

\usepackage{amsmath}
\usepackage{latexsym,amssymb}

\begin{document}
\title{Dynamics of blood cells  during a routine laboratory examination }

\author{Mesfin Asfaw  Taye}
\affiliation {West Los Angeles College, Science Division \\9000  Overland Ave, Culver City, CA 90230, USA}

\email{tayem@wlac.edu}

\begin{abstract} 
\section* {Abstract}
Centrifugation is a commonly performed laboratory procedure that helps to separate 
blood cells such as $RBCs$, $WBCs$, and platelets from plasma or serum.   Although centrifugation is a routine procedure in most medical laboratories,  the factors that affect the efficacy of the centrifugation process have never been studied analytically.  In this paper,   we examine the effect of the centrifugation time on the efficacy of the centrifugation process by studying the dynamics of the blood cells via the well-known Langevin equation or equivalently,  by solving the Fokker-Plank equation.  Our result depicts that the speed of the centrifuge
is one of the determinant factors concerning the efficacy of the centrifugation process. As the angular speed increases, the centrifugal force steps up and as result, the particles are forced to separate
from the plasma or serum. The room temperature also considerably affects the dynamics of analyse during centrifugation. Most importantly,  the generation of heat during centrifugation steps up the temperature within a centrifuge and as a result, not only the stability of the sample but also mobility of analyse is affected.    We show that as the centrifuge temperature steps up, the velocity of the cells as well as the displacement of the cell in the fluid decreases. We then study the dynamics of the whole blood during capillary action where in this case the blood flows upward in a narrow space without the assistance of external forces. Previous investigations show that the height that the fluid rises increases as the surface tension steps up.  The viscosity of the fluid also affects the capillary action but to date, the dependence of the height on viscosity has never been explored due to the lack of a mathematical correlation between the viscosity of blood and surface tension  \cite{aaa1}. In this work, we first examine the correlation between
surface tension and viscous friction via data fitting. Our result exhibits  that the viscosity of the blood increases linearly as the surface tension increases.  The  mathematical relation between the height and viscous friction is derived. It is shown that 
the height of the blood that rises in capillary  increases as the viscous friction steps up. As the temperature of the room steps up, the height also decreases. The  dependence of erythrocytes sedimentation rate
on surface tension is also studied.  The results obtained in this work show that 
the ESR increases as surface tension steps down
\end{abstract}
\pacs{Valid PACS appear here}
\maketitle
 
\section {Introduction}

Medical laboratory examinations are vital since these examinations help to diagnose any abnormalities and treat a  patient based on the observed results.  Particularly, blood tests are routinely performed to evaluate any abnormal conditions. Most of these blood works require sedimentation either via centrifugation or gravity. Often, the observed diagnostic test results are affected by external factors such as temperature. To understand the factors that affect the outcome of the routine examinations,   it is vital to explore the dynamics of the whole blood, erythrocytes (RBCs), leukocytes (WBCs), and thrombocytes (platelets). In this work, using physiological parameters, we study the dynamics of blood cells during routine lab exams. 

Particularly, centrifugation is one of the commonly performed laboratory procedures that help to separate  
blood cells such as $RBCs$, $WBCs$, and platelets from plasma or serum. When blood is first mixed with an anticoagulant and allowed to be centrifuged 
for a few minutes,  the blood cells sediments by leaving the plasma at the top. In the absence of anticoagulant, the blood clots, and when it is centrifuged, the blood cells sediment 
leaving the serum at the top. The serum is an ideal sample for diagnostic tests since it lacks leukocytes, erythrocytes, platelets, and other clotting factors.  Although centrifugation is a routine procedure in most medical laboratories,  the factors that affect the efficacy of the centrifugation process have never been studied analytically. As discussed in the work \cite{aaa1}, the centrifugation time, temperature, the length of the test tube, and the speed of the centrifuge 
are the determinant factors with regards to the efficacy of the centrifugation process. In this paper,  via an exact analytical solution,  we study the factors 
that affect the efficacy of the centrifugation process. 
First, we examine the effect of the centrifugation time on the efficacy of the centrifugation process.  Since blood cells are microscopic in size, their  dynamics can be modeled as a Brownian particle walking in a viscous medium.  As  blood is a highly viscous medium, the chance for the blood cells  to accelerate is negligible and the corresponding dynamics can be studied via Langevin equation or Fokker Planck equation \cite{aa7,aa8,aa9,aa10,aa11,aa16,aa17}. Solving the Fokker Planck equation analytically, we explore how  the dynamics of blood cells  behave as a function of the model parameters. Because   our study is performed by considering real physiological parameters, the results obtained in this work non only agree with the experimental observations but also help to understand most hematological experiments that are conducted in vitro.   
In a medical laboratory, the standard test tube has a length of $L=150$ $mm$ and the cells separate from the plasma or serum at the result of the centrifugal force $f=Nm\omega^2 r $ where $N$  denotes the number of cells that form a cluster while $m$ designates the mass of $RBC$ or platelets.  Since the $WBCs$ are heavy, they move way before $RBCs$, and as an approximation one can disregard their dynamics.

Our result depicts that the speed of the centrifuge
is one of the determinant factors concerning the efficacy of the centrifugation process. As the angular speed increases, the centrifugal force steps up and as result, the particles are forced to separate
from the plasma or serum. 
Depending on the size and fragility of the sample,  the centrifugation speed should be adjected.  To increase the efficacy of the centrifugation process,  as the size of the particle decreases,  the centrifugation speed should step up. The length of the test tube affects the effectiveness of the centrifugation process. As the length of the test, tube steps up, the centrifugal force increases.
The room temperature considerably affects the dynamics of analyse during centrifugation. Most importantly,  the generation of heat during centrifugation steps up the temperature within a centrifuge and as a result, not only the stability of the sample but also mobility of analyse is affected.  The effect of temperature near the test tube causes difficulties in the current experimental procedure by inducing additional false-positive results.  In this regard, developing a mathematical model and exploring the model system using the powerful tools of statistical mechanics provides insight as well as guidance regarding the dynamics of RBCs in vitro.  Our result shows that as the centrifuge temperature steps up, the velocity of the cells as well as the average distance of the cell in the fluid decreases. This effect of temperature can be corrected by increasing the centrifugation time. However,  a longer centrifugation time might lead to a considerable amount of heat generation. This, in turn,  may cause the red blood cells to lyse. Prolonged centrifugation at high speed also causes structural damage to the cells and as a result, hemolysis occurs.  On the contrary, low-speed centrifugation leads to insufficient separation of plasma and serum from cellular blood components. 

Our analysis also indicates that, when the RBC forms rouleaux (as $N$ increases), the velocity of the cells increases. This is because as $N$ steps up, the centrifugal force increases.  This also implies since the cells in serum form aggregates due to clotting factors, they need less centrifugation time than plasma.  
As shown in Fig. 1, the size of WBC is considerably large in comparison with the RBC. Since the platelet has the smallest size, its velocity and displacement along the test tube are significantly small.  
As a result, the  RBCs separate far more than platelets showing that to avoid the effect of clotting factors,  one has to increase the centrifugation time.  
\begin{figure} 
 \includegraphics[width=6cm]{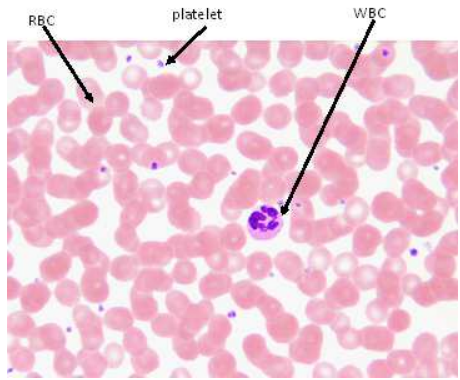}
\caption{ A blood smear that shows the composition of RBCs, WBCs, and platelets  \cite{aaa5}.}
\end{figure}

Furthermore, we study the dynamics of the whole blood during capillary action where in this case the blood flows upward in narrow spaces without the assistance of external forces. Previous investigations show that the height that the fluid rises increases as the surface tension steps up.  The viscosity of the fluid also affects the capillary action but to date, the dependence of the height on viscosity has never been explored due to the lack of a mathematical correlation between the viscosity of blood and surface tension  \cite{aaa2}. In the past, for non-Newtonian fluids, Pelofsky \cite{aaa3} has studied the relation between surface tension and viscosity.  His empirical equation depicts that the surface tension increases as the viscosity steps up. In this work, we first examine the correlation between
surface tension and viscous friction via data fitting. We show  that the viscosity of the blood increases as the surface tension increases.  The  mathematical relation between the height and viscous friction is also derived. It is shown that 
the height that the blood rises increases as the viscous friction steps up. As the temperature of the room steps up, the height also decreases.  

Moreover, in this work, we also explored, the dependence of the erythrocyte sedimentation rate (ESR) on model parameters.   As discussed in our recent paper  \cite{aaaa3},   the erythrocyte sedimentation rate (ESR) often measures how fast a  blood sample sediments along a test tube in one hour in a clinical laboratory. This analysis is performed by mixing whole blood with an anticoagulant. The blood is placed in an upright  Wintrobe or Westergren tube and allowed to sediments for an hour. The normal values of the ESR varies from 0-3mm/hr for men and 0-7 mm/hr for women \cite{aa2,aa3,aa4}.  High ESR   is associated with diseases that cause inflammation.  In the case of inflammatory disease, the blood level of fibrinogen becomes too high \cite{aa4, aa6}. The presence of fibrinogen forces the RBCs to stick each other and as a result, they form aggregates of RBC called rouleaux. As the mass of the rouleaux increases, the weight of the rouleaux dominates the vicious friction and as a result, the RBCs start to precipitate.   The temperature of the laboratory (blood sample)  also significantly affects the test result \cite{s4}. As the temperature of the sample steps up, the ESR increases.  In the past,  a mathematical model was developed by Sharma $.et
$. $al$ to study the effect of blood concentration on the erythrocyte sedimentation rate \cite{s1}. Later the effect of concentration of nutrients on the red blood cell sedimentation rate was investigated in the work \cite{s2}.  More recently,  the sedimentation rate of RBC  was explored via a model that uses Caputo fraction derivative \cite{s3}. The theoretical work obtained in this work was compared with the sedimentation rate that was analyzed experimentally.  All of these experimental and theoretical works exposed the factors that affect the sedimentation rate of RBCs.   In this paper, extending our recent work \cite{aaaa3}, we study how surface tension as well at the tilt in test tube angle affects the ESR.

The rest of the paper is organized as follows. In section II, we present the model system. In section III,  we study the factors 
that affect the efficacy of the centrifugation process.  The dynamics of the whole blood during capillary action  is studied in section IV. In section V, the dependence of
erythrocytes sedimentation rate on model parameters is studied. Section VI deals with summary and conclusion.

\section{The model}  

Since RBC is microscopic in size, its dynamics (in vitro)  can be  modeled as a Brownian particle that undergoes a biased random walk  on one-dimensional  test tube.  In the routine hematology test, the erythrocyte  dynamics is also affected by 
gravitational  force 
\begin{equation}
  f=Nmg 
\end{equation}
and centrifugal force 
\begin{equation}
  f=Nm\omega^2 r 
\end{equation}
where  $g=9.8m/s^{2}$ is the gravitational acceleration. $N$ denotes the number of blood cells that forms rouleaux. $\omega$ denotes the angular speed  of the centrifuge (see Fig.2) and $r$ designates the radius of the shaft. The  speed of clinical centrifuge  varies  from $ 200$ rpm ( $ 21 rad/s$) and $21000$ rpm ( $ 2198 rad/s$). 
 The mass of the red blood cells  $m=27X10^{-15}$ kg. Since platelets are only $20$ percent of the size of $RBC$, we infer the mass of the platelets to be $m=5.2 X10^{-15}$ kg. The average size of  RBC and platelets  is given as $r'=4X10^{-6}$  and  $r'=7.5 X10^{-7}$ meter, repectively. The normal value of RBC  on average  varies from $5X10^6-6X10^6/mm^3$ \cite{aa12,aa13,aa14}. 
\begin{figure} 
 \includegraphics[width=6cm]{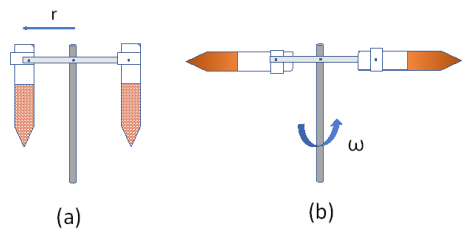}
\caption{A centrifuge in a clinical  laboratory is used  to separate the   
blood cells such as $RBCs$, $WBCs$, and platelets from the plasma or serum. Here $r$ designates the radius of the shaft.   }
\end{figure}

{\it Underdamped case :\textemdash} 
The dynamics  of the RBC in vitro is governed   by  the well-known Langevin equation 
\begin{eqnarray}
{dV\over dt}&=& -\gamma V-f + \sqrt{2k_{B}\gamma T}\xi(t).
\end{eqnarray}
 The random noise $\xi(t)$ is assumed to be Gaussian white noise satisfying the relations 
$\left\langle  \xi(t) \right\rangle =0$ and $\left\langle \xi(t)  \xi(t') \right\rangle=\delta(t-t')$.  The viscous friction  $\gamma$  and  $T$  are assumed to be spatially invariant along with the medium.
For  a non-Newtonian  fluid such blood, it is reasonable  to assume  that   
when  the  temperature of the blood sample  increases by $1$ degree Celsius, its viscosity  steps down by $2$ percent \cite{aa15} as 
 $\gamma'  =    B-{2B\over 100}(T-T^R)$
where  $B=4X10^{-3} kg/ms$ is the dynamical  viscosity of blood at a room temperature ($T^R=20$ degree Celsius) and $T$ is the temperature \cite{aa16}.  On the other hand,  from Stokes's  theorem, the viscosity $\gamma=6r' \pi \gamma'$. Here $k_{B}=1.38 X 10^{-23} m^{2} kg s^{-2} K^{-1}$ is the  Boltzmann constant. 

Alternatively, Eq. (3) can be rewritten  as a   Fokker-Plank equation  
\begin{eqnarray}
{\partial P\over \partial t}&=&-{\partial (vP) \over \partial x}-{1 \over m}{\partial(fP) \over \partial v}+ \nonumber \\
&&{\gamma \over m}{\partial (vP) \over \partial v}+{\gamma T \over m^2}{\partial^2 P \over \partial v^2}
\end{eqnarray}
where $P(x,v,t)$ is the probability  of finding the particle at particular position $x$, velocity $v$ and time $t$.

For convenience, Eq. (4)  can be rearranged     as 
\begin{eqnarray}
{\partial P\over \partial t}&=& -(k+{\partial J' \over \partial v} )
\end{eqnarray}
where 
\begin{eqnarray}
k=v{\partial P \over \partial x} ={\partial J \over \partial x}
\end{eqnarray}
and 
\begin{eqnarray}
J'= -{\gamma (x,t)\over m}vP+{1\over m}(U'P) -{\gamma  T\over m^2}{\partial P\over \partial v}.
\end{eqnarray}
From Eqs. (6) and (7), one gets 
\begin{eqnarray}
{\partial P \over \partial v}= -{m^{2}J'\over \gamma T}+{mU'P\over \gamma  T}-{m vP\over T}
\end{eqnarray}
and  
\begin{eqnarray}
{\partial P \over \partial x}= {k\over v}.
\end{eqnarray}

 After some algebra, the expression for  
the probability distribution $ P(v,t) $  is given as 
\begin{eqnarray}
P(v,t)&=&{e^{-{m({-(1-e^{-{\gamma t \over m}})f\over \gamma }+v)^2\over 2(1-e^{-{2\gamma t \over m}})T}}\sqrt{{m\over (1-e^{-{2\gamma t \over m})}t}}\over \sqrt{ 2\pi}}.
\end{eqnarray}
 
The velocity of the cell can be evaluated as 
\begin{eqnarray}
V(t)&=&\int_{0}^{\infty}P(v',t)v'dv' \nonumber \\
&=&\left({1-e^{-{\gamma t\over m}}\over \gamma}\right)f.
\end{eqnarray}
Once the velocity of the cell is calculated, the position of the cell is then given by
\begin{eqnarray}
x&=&\int_{0}^{t}V(t')dt'.
\end{eqnarray}
Here one should note that the solutions  for  overdamped and underdamped cases are not new 
since such cases are presented by a well-known Ornstein-Uhlenbeck process.

{\it Overdamped case:\textemdash} Blood is a highly viscous medium, as the result, the chance for the RBC to accelerate is negligible. One can then  neglect the inertia effect  and the corresponding dynamics can be studied via the Langevin equation 
\begin{eqnarray}
\gamma{dx\over dt}&=& -f + \sqrt{2k_{B}\gamma T}\xi(t).
\end{eqnarray}

One can also write Eq. (13) as  a Fokker-Plank equation 
\begin{equation}
{\partial P(x,t)\over \partial t}={\partial\over  \partial x}
\left[{f\over \gamma}P(x,t)+{\partial \over \partial x}\left({k_{B}T\over \gamma}P(x,t)\right)\right]
\end{equation}
where $P(x,t)$ is the probability density of finding the particle (the cell) at position $x$ and  time $t$.

 To calculate the desired thermodynamic quantity, let us first find the probability distribution. After imposing  a periodic boundary condition $P(0,t)=P(L,t)$, we solve Eq. (14). After some algebra, one finds the   probability distribution  as 
\begin{eqnarray}
P(x,t)&=&\sum_{n=0}^\infty \cos[{n\pi \over L}(x+t {f\over \gamma})]e^{-({n\pi  \over L})^2t{k_{B}T\over \gamma}}
\end{eqnarray}
where  $T$ is the temperature of the medium. For detailed mathematical analysis, please refer to my previous work  \cite{aa16}. Next we use Eq. (15)  to find the  particle current, the velocity  as well as the position of the particle.
 The particle current is then given by 
\begin{eqnarray}
J(x,t)&=&-\left[f P(x,t) + k_{B}T{\partial P(x,t) \over \partial x}\right].
\end{eqnarray}
After substituting  $P(x,t)$ shown in Eq. (15), one can find 
the velocity of the cells at any time as 
\begin{eqnarray}
V(x,t)&=&\int_{0}^{x}J(x',t)dx'
\end{eqnarray}
while the position of the cells can be found  via
\begin{eqnarray}
x&=&\int_{0}^{x}P(x',t)x'dx'.
\end{eqnarray}

The fact that blood is a highly viscous medium (since $\gamma$ is considerably high), the numerical value of velocity calculated via Eq. (11) is approximately the same as the velocity calculated via Eq. (17).
At steady state (in long time limit), the velocity (Eqs. (11) and (17) ) approach $V=f/\gamma$. One should also note that the diffusion constant for the model system is given by  $D={k_BT\over \gamma}$. This equation is valid when viscous friction is temperature-dependent showing that the effect of temperature on the mobility of the cells 
is significant. When temperature increases, the viscous friction gets attenuated and as a 
result the diffusibility of the particle increases. Various experimental studies also showed
that the viscosity of the medium tends to decrease as the temperature of the medium 
increases \cite{aa1}. This is because increasing the temperature steps up the speed of the 
molecules, and this in turn creates a reduction in the interaction time between
neighboring molecules.  As a result, the intermolecular force between the molecules 
decreases, and hence the magnitude of the viscous friction decreases.

\section{The dynamics of  blood cells  during centrifugation  }

As a common standard practice, in a clinical laboratory,   centrifugation is vital to separate the   
blood cells such as $RBCs$, $WBCs$, and platelets from the plasma or serum. When blood is first mixed with an anticoagulant and allowed to be centrifuged 
for a few minutes,   the blood cells separate from the plasma. In the absence of anticoagulant, the blood clots, and when it is centrifuged, the blood cells segregate from the serum. In this section, we study   the factors that affect the efficacy of the centrifugation process  analytically. We show that the centrifugation time, temperature, the length of the test tube, and the speed of the centrifuge 
are the determinant factors with regards to the efficacy of the centrifugation process.

First, let us examine the effect of the centrifugation time on the efficacy of the centrifugation process. This can be investigated by tracking the dynamics of the blood cells during centrifugation. The dynamics of the cells in vitro is governed by the well-known Langevin equations (3) or (13).  Equivalently,  by solving Fokker-
Plank equations (4) or (14), the information regarding the mobility of the RBCs can be extracted.
In a medical laboratory, the standard test tube has a length of $L=150$ $mm$ and the cells separate from the plasma or serum at the result of the centrifugal force $f=Nm\omega^2 r $ where $N$  denotes the number of cells that form a cluster while $m$ designates the mass of $RBC$ or platelets.  Since the $WBCs$ are heavy, they move way before $RBCs$, and as an approximation one can disregard their dynamics.

Exploiting Eqs. (11) or (17)  one can see that the velocity of the RBC steps up and saturates to a constant value (see Fig. 3a). In the figure,  we fix the parameters as $N=1$ (at $20$ degree celsius), $\omega=300 rad/s$, $r=0.5m$ and  $L=0.15m$. Via Eqs. (12) or (18), one can track the position of RBC along the test tube during the centrifugation processes. As depicted in Fig. 4a, the particle walks away towards the bottom of the test tube as time steps up. Unlike serum, plasma contains platelets and this indicates that more configuration time is needed for plasma since platelets are small in size.  Fig. 4a is plotted by fixing $N=1$,  $T=20$ degree celsius,  $\omega=300 rad/s$, $r=0.5m$ and  $L=0.15m$. The three-dimensional plot depicted in Fig. 5 also confirms that as the centrifugation time steps up, the cells segregate and move towards the bottom of the test tube.

\begin{figure}[ht]
\centering
{
    \includegraphics[width=6cm]{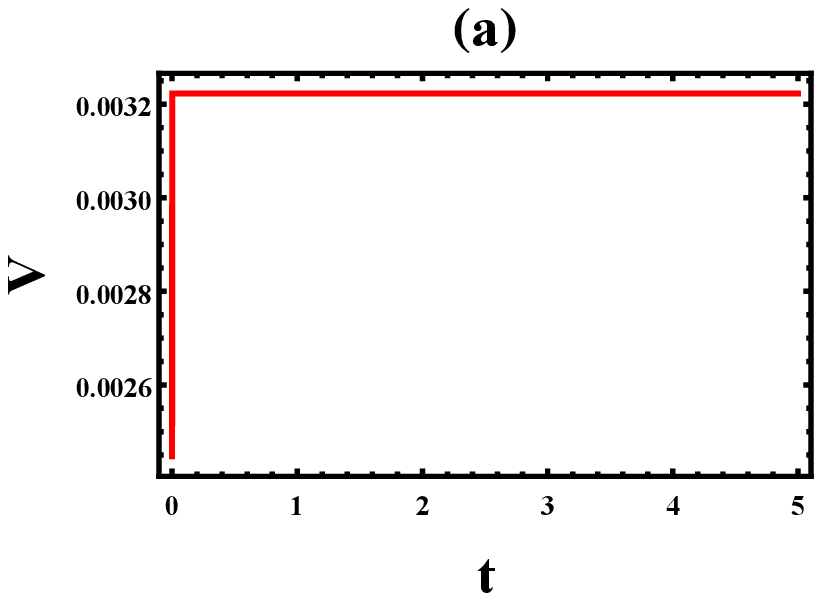}
}
\hspace{1cm}
{
    \includegraphics[width=6cm]{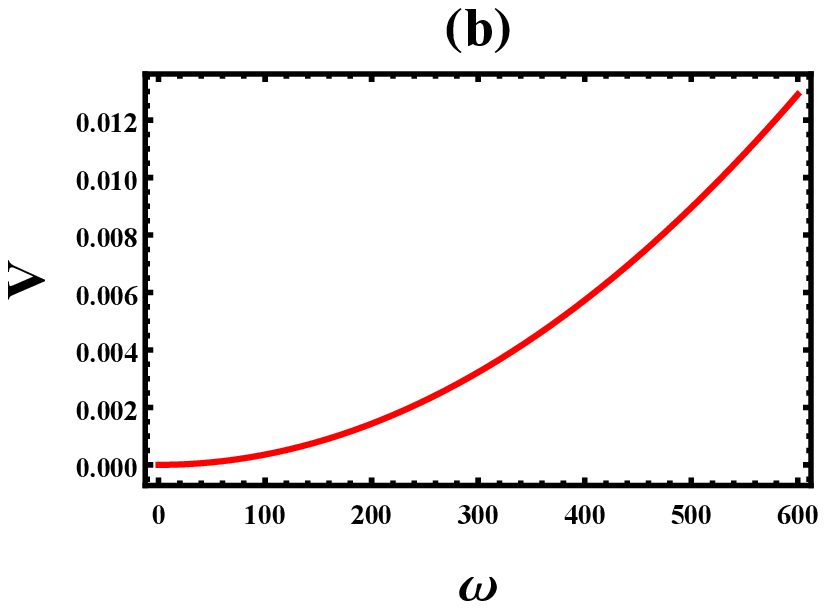}
}
\caption{ (Color online)(a) The velocity ($V (m/s)$) of RBC as a function of time $t$ (in seconds) for a single RBC $N=1$,  $T=20$ degree celsius, $\omega=300 rad/s$, $r=0.5m$ and  $L=0.15m$ .
 (b)  The velocity ($V(m/s)$) of RBC as a function of  angular velocity $\omega$ (in rad/s)  for fixed values of   $T=20$ degree celsius,  $t=180 s$, $N=1$, $r=0.5m$ and  $L=0.15m$. } 
\label{fig:sub} 
\end{figure}

The speed of the centrifuge
also affects the efficacy of the centrifugation process.  As the angular speed increases  (see Eq. (2) ), the centrifugal force steps up and as result, the particles are forced to separate
from the plasma or serum. 
Depending on the size and fragility of the analyse,  the centrifugation speed should be adjusted.  To increase the efficacy of the centrifugation process,  as the size of the particle decreases,  the centrifugation speed should step up.
The dependence of the speed of the particle on  angular speed  is  explored. As depicted in Fig. 3b, the velocity of the RBC steps up monotonously as the angular speed steps up.  In the figure  we fix the parameters as $N=1$ (at $20$ degree celsius), $\omega=300 rad/s$, $r=0.5m$ and  $L=0.15m$. One can also track the position of RBC along the test tube as a function of angular speed. 
 As shown in Fig. 4b, the particle moves towards the bottom  of the test tube as angular speed steps up. The figure is plotted by fixing $N=1$,  $T=20$ degree celsius,  $\omega=300 rad/s$, $r=0.5m$ and  $L=0.15m$.
\begin{figure}[ht]
\centering
{
    \includegraphics[width=6cm]{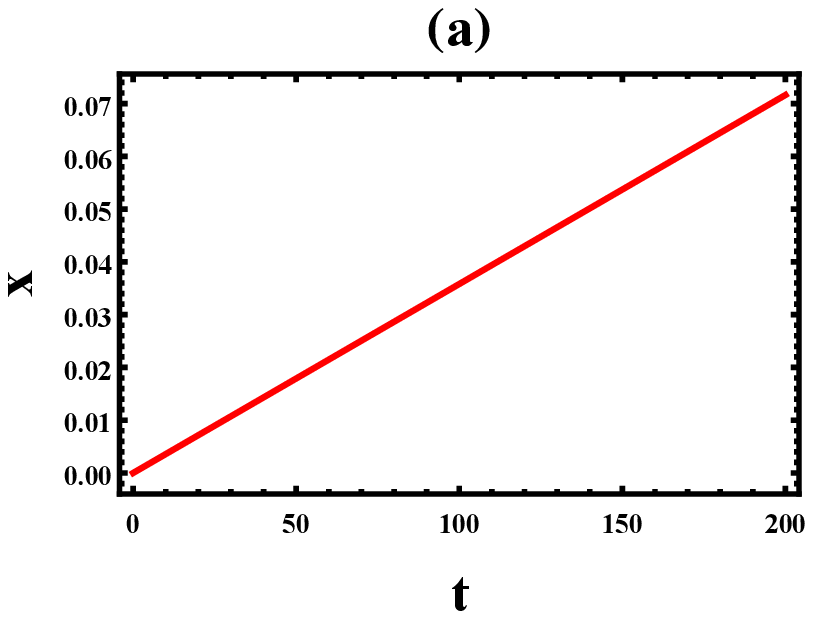}
}
\hspace{1cm}
{
    \includegraphics[width=6cm]{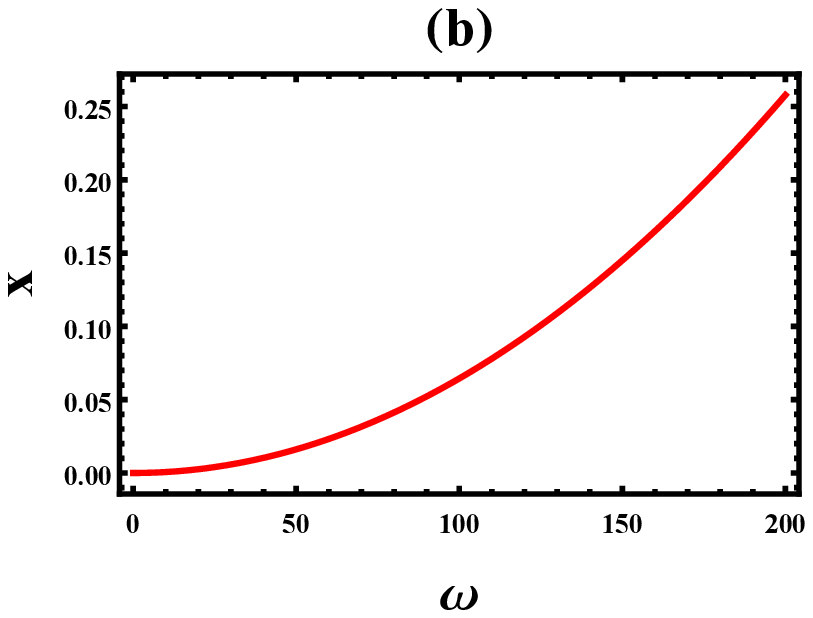}
}
\caption{ (Color online)(a) The sedimentation displacement ($x (m)$) of RBC as a function of time $t$ (in seconds) for fixed  $\omega=100 rad/s$.
 (b)  The sedimentation distance ($x (m)$) of RBC as a function of  $\omega$ (in rad/s) for a given $t=180 s$. For both figures, we fix $T=20$ degree celsius,  $r=0.5m$ and  $L=0.15m$.}
\label{fig:sub} 
\end{figure}
\begin{figure}[ht]
\centering
{
    \includegraphics[width=6cm]{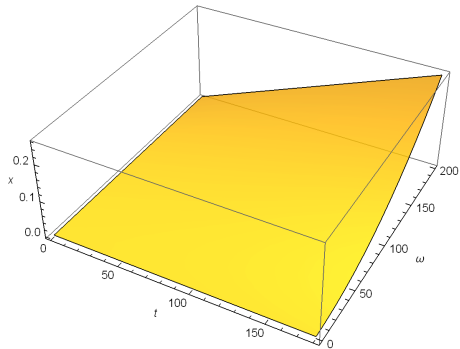}
}
\caption{ (Color online) The displacement   $x$   as a function of time $t$ and  $\omega$ for  fixed values of $T=20$ degree celsius, $r=0.5m$ and  $L=0.15m$.} 
\label{fig:sub} 
\end{figure}

The length of the test tube affects the effectiveness of the centrifugation process. As the length of the test tube steps, the centrifugal force increases.
The room temperature considerably affects the dynamics of analyse during centrifugation. Most importantly,  the generation of heat during centrifugation steps up the temperature within a centrifuge and as a result, not only the stability of the sample but also mobility of analyse is affected.  The effect of temperature near the test tube is unavoidable and causes difficulties in the current experimental procedure by inducing additional false-positive results.  In this regard, developing a mathematical model and exploring the model system using the powerful tools of statistical mechanics provides insight as well as guidance regarding the dynamics of RBCs in vitro.  

The role of temperature on the mobility of the cells during centrifugation can be appreciated by analyzing  Eq. (11). Substituting Eq. (2) into Eq. (11), one gets 
\begin{eqnarray}
V(t)&=&Nm\omega^2 r\left({1-e^{-{\gamma t\over m}}\over \gamma}\right).
\end{eqnarray}
The viscosity $\gamma$ of the blood is the function temperature  as $\gamma=6r' \pi \gamma' (B-{2B\over 100}(T-T^R))$ where $r'$ is the radius of the cells. This implies the velocity $V(t)$ and position $x$ are also the function of temperature. From Eq. (19) one can see that as the centrifuge temperature increases, the velocity of the cells as well as the displacement of the cell in the fluid decreases. As discussed before, the effect of temperature can be corrected by increasing the centrifugation time. However,  a longer centrifugation time might lead to a considerable amount of heat generation. This, in turn,  may cause the red blood cells to lyse. Prolonged centrifugation at high speed also causes structural damage to the cells and as a result, hemolysis occurs.  On the contrary, low-speed centrifugation leads to insufficient separation of plasma and serum from cellular blood components.  By analyzing  Eq. (19), one can also deduce that when the RBC form rouleaux (as $N$ increases), the velocity and the position of the particle step. This is because as $N$ steps up, the centrifugal force increases.  This also implies since the RBC in serum forms aggregates due to clotting factors, it needs less centrifugation time than plasma.  
As shown in Fig. 1, the size of WBC is considerably large in comparison with the RBC. Since the platelet has the smallest size, its velocity and displacement along the text tube are significantly small. As depicted in Fig. 6a,  the velocity for platelets is considerably lower than red blood cells due to the small size of platelets.  Moreover, as shown in Fig. 6b, the  RBC  moves far more than platelets showing that to avoid the effect of clotting factors,  one has to increase the centrifugation time.  
\begin{figure}[ht]
\centering
{
    \includegraphics[width=6cm]{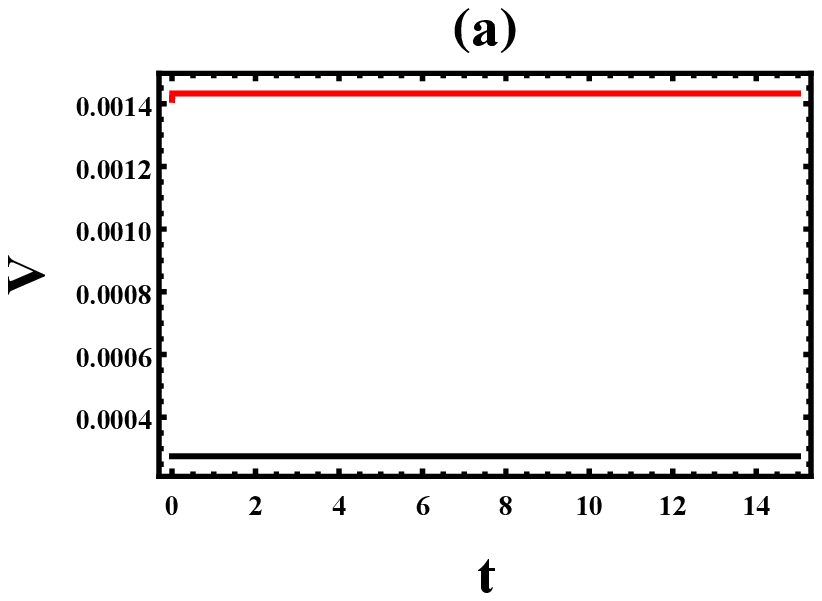}
}
\hspace{1cm}
{
    \includegraphics[width=6cm]{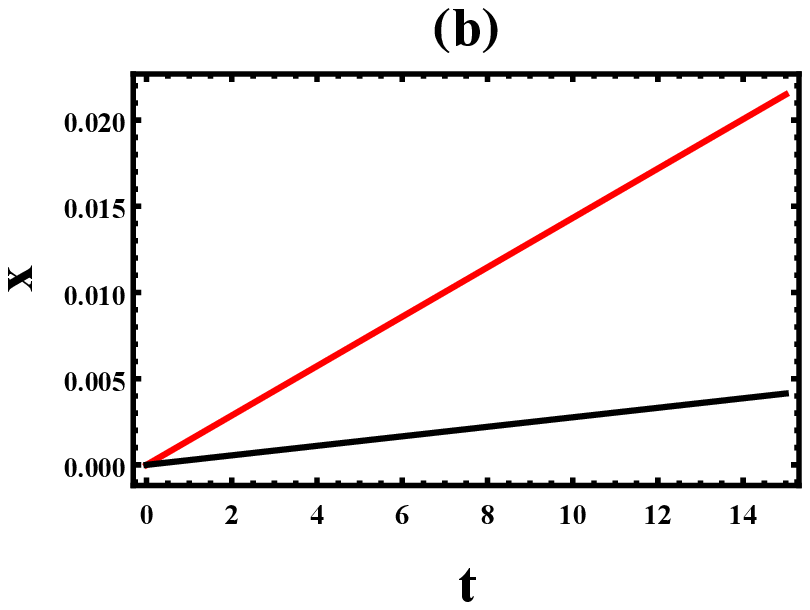}
}
\caption{ (Color online) (a) The sedimentation velocity  $V$ of RBC (red line) and platelet (black line) as a function of time $t$. 
 (b)  The displacement    of RBC (red line) and platelet (black line) as a function of time $t$. In the figures, we fix   $T=20$ degree celsius, $\omega=200 rad/s$, $r=0.5m$ and  $L=0.15m$.}
\label{fig:sub} 
\end{figure}

\section{The dynamics of whole blood during capillary action }

In this section, we study the dynamics of the whole blood during capillary action where in this case the blood flows upward in narrow spaces without the assistance of external forces. Previous investigations show that the height that the fluid rises increases as the surface tension steps up.  The viscosity of the fluid also affects the capillary action but to date, the dependence of the height on viscosity has never been explored due to the lack of mathematical correlation between the viscosity of blood and surface tension  \cite{aaa2}. In the past, for non-Newtonian fluids, Pelofsky \cite{aaa3} has studied the relation between surface tension and viscosity.  His empirical equation depicts that the surface tension increases as the viscosity steps up. In this work, we first analyzed the correlation between
surface tension and viscous friction via data fitting. 
\begin{table}[h!]
\centering
\begin{center}
\begin{tabular}{ ||c|c|c|| } 
 \hline
 $s$&$\gamma'$&$ T (c^{o})$ \\
			0.06064& 0.004&20 \\
			0.060167& 0.00392& 21 \\
			0.059694& 0.00384&  22 \\
			0.059221& 0.00376& 23 \\
			0.058748& 0.00368& 24 \\
			0.058275&  0.0036& 25 \\
			0.057802& 0.00352& 26 \\
			0.057329& 0.00344& 27  \\
			0.056856& 0.00336& 28 \\ 
			0.056383& 0.00328& 29 \\
			0.05591&0.0032& 30 \\
			0.055437& 0.00312& 31 \\
			0.054964& 0.00304& 32 \\
			0.054491& 0.00296& 33 \\
			0.054018& 0.00288& 34 \\
			0.053545&  0.0028& 35 \\
			0.053072& 0.00272& 36 \\ 
			0.052599& 0.00264&  37 \\
			0.052126& 0.00256& 38 \\
			0.051653& 0.00248& 39 \\
			0.05118&0.0024& 40 \\		
			 \hline
\end{tabular}
\end{center}
\caption{Data that show the dependence of surface tension $S$ and viscous friction  $\gamma'$ on $T$.}
\label{table:1}
\end{table}

The surface tension of blood  is a function of temperature  $T$  and it is given by \cite{aaa2} 
\begin{eqnarray}
S=(-0.473 T+70.105 )X 10^{-3} N/m.
\end{eqnarray}
As discussed before, the dynamical viscous friction of blood depends on  the temperature as 
\begin{eqnarray}
\gamma'  =    B-{2B\over 100}(T-T^R)
\end{eqnarray}
where  $B=4X10^{-3} kg/ms$ is the dynamical  viscosity of blood at a room temperature ($T^R=20$ degree Celsius) and $T$ is the temperature \cite{aa16}. The correlation between surface tension can be inferred from data fitting.  Via Eqs. (20) and (21),  the dependence of $S$ and $\gamma'$ on $T$ is explored as shown in Table I. Since the experiments show the surface tension  $S$  to have functional dependence on  $\gamma'$,    by fitting  the data depicted in Table I, one finds
\begin{eqnarray}
S&=&0.03699 + 5.9125 ( B-{2B\over 100}(T-T^R)) \nonumber \\
&=&0.03699 + 5.9125 \gamma'.
\end{eqnarray}

Furthermore, the surface tension is responsible for a liquid to rise up in the test tube against the downward gravitational force.  As a result of the capillary action, the fluid  rises   to the height 
\begin{eqnarray}
h={2S cos(\theta) \over \rho g z}
\end{eqnarray}
where $S$ is the surface tension of the blood, $\rho=1060 kg/m^3$ is the density of the fluid, $g=9.8 {m/s^2}$ is the gravitational acceleration and $ z$ is the radius of the test tube. $\theta$ is the contact angle between the blood and the test tube while $h$ is the height that the blood rises up in the test tube. For the case where 
$\theta=0$,  we rewrite Eq. (23) as
\begin{eqnarray}
s={h\rho g z \over 2}.
\end{eqnarray}
From Eqs. (22) and (24), after  some algebra we get 
\begin{eqnarray}
h= {0.074+11.824\gamma' \over  g  z \rho}.
\end{eqnarray}
Equation (25)  exhibits that the height $h$ is a function of test tube radius $z$ and the density of the fluid $\rho$. In a clinical laboratory, the  Haematocrit tube has a height of $ 75mm$ and   inner diameter $z=0.4mm$. Substituting these parameters in Eq. (25), the height that the blood rises is found to be $58.4mm$ at $20$ degree Celsius.  
The height $h$ that the blood rises in the capillary tube is also sensitive to temperature.  Via Eqs. (21) and (25), one can see that as the room temperature steps up, $h$ decreases  (see Fig. 6).
\begin{figure}[ht]
\centering
{
    \includegraphics[width=6cm]{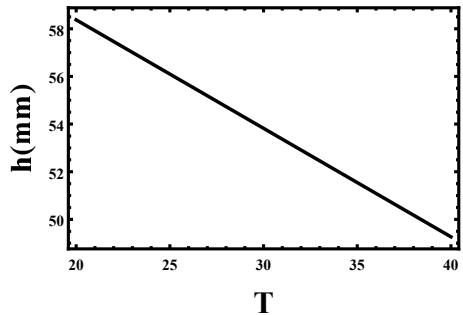}
}
\caption{ (Color online) The height $h$ that the blood rises in the capillary tube as a function of temperature for fixed-parameters  $z=0.4mm$ and  $\rho=1060 kg/m^3$. } 
\label{fig:sub} 
\end{figure}

The above analysis points that the increase in the
room temperature results
in a false positive or negative result. As shown in Fig. 7,   up to
8mm sedimentation  can be observed
when the temperature of the room varies from 20 to 40
degree celsius. Note that the shape of RBC, plasma
viscosity, and inclination of the test tube also affect the
magnitude of surface tension. In anemic patients, a low hematocrit level is observed, and consequently the surface tension or 
viscosity of the blood decreases which in turn results in lower $h$. Exessive use of 
 anticoagulants results in lower $h$. This is because adding too much anticoagulant
decreases the viscosity or surface tension of the blood. 

\section{The role  of  surface tension on Erythrocytes sedimentation rate and  the dynamics of  blood cells  during centrifugation  }

In this section, we explore  how the surface tension of the blood affects the mobility of RBC by solving the model system analytically.

{\it Case 1:  The effect  of surface tension on the erythrocyte sedimentation rate.\textemdash}
The erythrocyte sedimentation rate (ESR) often measures how fast a  blood sample sediments along a test tube in one hour in a clinical laboratory as shown in Fig. 11. By mixing whole blood with an anticoagulant, the blood is placed in an upright  Wintrobe or Westergren tube and allowed to sediments for an hour.  In the  Westergren method, the test tube is $200 mm$  long while in the Wintrobe method the tube is only $100 mm $ long. 
To  investigate the effect of surface tension, let us rewrite Eq. (22) as 
\begin{eqnarray}
\gamma'={(S-0.03699)\over  5.9125}
\end{eqnarray}
and  after some algebra  we get 
\begin{eqnarray}
\gamma=6r'\pi{(S-0.03699)\over  5.9125}.
\end{eqnarray}
 Substituting Eqs. (1) and (27) in Eq. (11),  one gets
\begin{eqnarray}
V(t)&=&Nmg\left({1-e^{-{6r'\pi{(S-0.03699)\over  5.9125}t\over m}}\over 6r'\pi{(S-0.03699)\over  5.9125}}\right).
\end{eqnarray}
The position of the cell is then given by
\begin{eqnarray}
x&=&\int_{0}^{t}V(t')dt'.
\end{eqnarray}

From Eqs. (28) and (29), it is evident that as the surface tension steps up,  the velocity and displacement of the cells step down.  Since the surface tension (Eq. 20) depends on temperature, the dynamics of the cells are also affected by the room temperature. Exploring  Eqs. (28) and (29), one can see that the velocity of the cell is significantly affected by the temperature of the room, and the size of the particle as shown in Fig. 8a. As the temperature steps up, the velocity increases.  The same figure also depicts that the RBC precipitates faster than platelets which is reasonable since platelets are small in size.  The plot of the position of the cell as a function of time is depicted in Fig. 8a.  The figure exhibits that the RBC has a  faster sedimentation rate  than platelets. 
\begin{figure}[ht]
\centering
{
    \includegraphics[width=6cm]{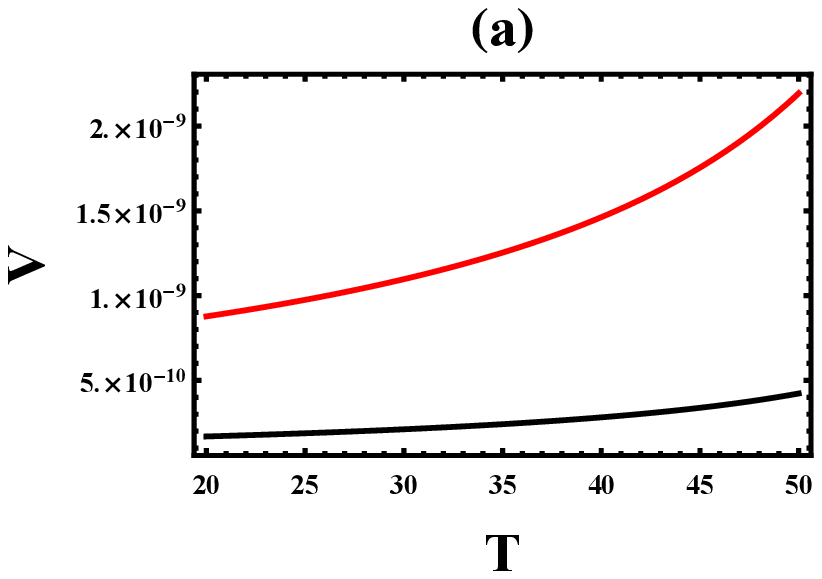}
}
\hspace{1cm}
{
    \includegraphics[width=6cm]{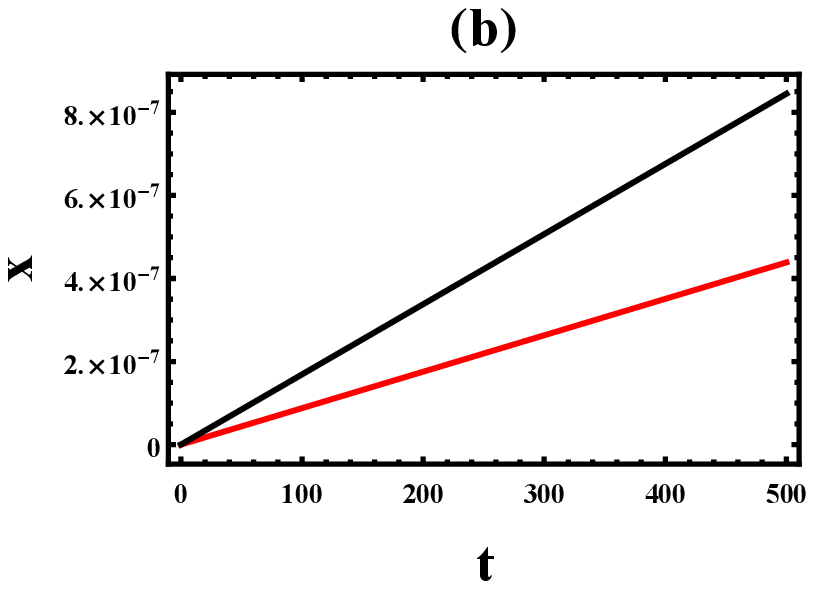}
}
\caption{ (Color online)(a) The velocity ($V$) of cell as a function of temperature $T$  for a single cell $N=1$ and $t=1000 s$.  (b) The displacement  of the cell as a function of time $T$  for a single cell $N=1$ and $T=20 $ degree celsius. In the figures, the red and black lines indicate the plot for the position of RBC and platelets,    respectively.}
\label{fig:sub} 
\end{figure}

From Eqs.  (28) and (29),  one can see that as  the RBC forms rouleaux (as $N$ increases)
 the sedimentation rate increases (see Fig. 9). 
Figure 9 depicts the plot of ESR as a function of the number of RBCs (at $20$ degree celsius). As shown in the figure, the ESR increases as $N$  steps up. 
\begin{figure}[ht]
\centering
{
    \includegraphics[width=6cm]{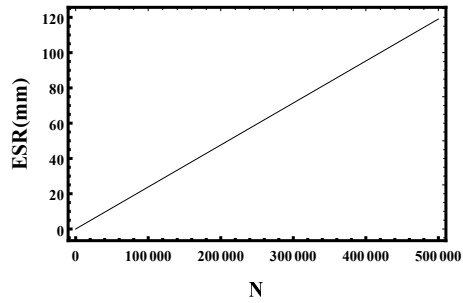}
}
\caption{ (Color online) Erythrocyte sedimentation  rate in one hour at $20$ degree celsius. The ESR steps up as the number of red blood cells that form rouleaux  increases. } 
\label{fig:sub} 
\end{figure}

 The background temperature of the fluid  also affects the viscous friction of the fluid. As temperature increases, the viscous friction decreases, and on the contrary, 
 the diffusibility of the cells increases as depicted in the works \cite{aa1, aa8}. For highly viscous fluid such as blood, when the temperature of the blood sample increases by $1$ degree celsius, its viscosity steps down by $2$ percent.  Exploiting Eqs. (28) and    (29),  the dependence of the erythrocyte sedimentation rate as a function of temperature (in degree celsius) is depicted in Figure 10.  In the figure, the number of RBCs that form clusters is fixed as $N=15X10^3$, $N=1X10^4$, and $N=5X10^3$ from top to bottom, respectively. The figure depicts that the ESR steps up as the number of red blood cells ($N$) that form rouleaux increases as well as when the temperature of the room steps up.
The same figure shows that up to 3mm/hr sedimentation rate difference can be observed when the temperature of the room varies from 20 to 45 degree celsius.  
\begin{figure}[ht]
\centering
{
    \includegraphics[width=6cm]{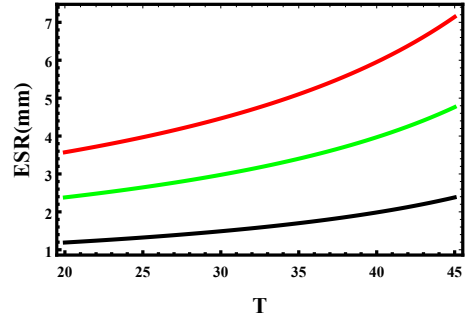}
}
\caption{ (Color online) Erythrocyte sedimentation rate in one hour as a function of temperature in degree celsius. The number of RBCs that form clusters is fixed as $N=15X10^3$, $N=1X10^4$, and $N=5X10^3$ from top to bottom, respectively. 
The ESR steps up as the number of red blood cells ($N$) that form rouleaux increases as well as when the temperature of the room steps up.} 
\label{fig:sub} 
\end{figure}

\begin{figure} 
 \includegraphics[width=6cm]{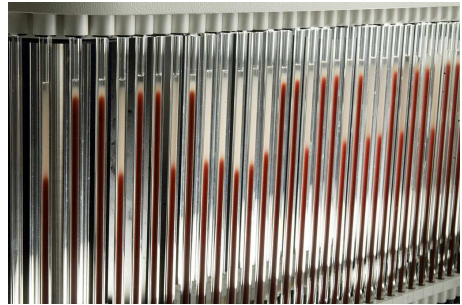}
\caption{  Figure that shows the sedimentation rate of  RBC  on Westergren pipet \cite{aa5}. 
This hematology test is performed by mixing whole blood with an anticoagulant.  The blood is then placed in an upright  Wintrobe or Westergren tube. The sedimentation rate of the red blood cells is measured in millimeters (mm) at the end of one hour. }
\end{figure}

{\it Case 2:  The role  of surface tension on  the dynamics of  blood cells  during centrifugation  .\textemdash} As blood cells are microscopic, they undergo a random motion in vitro when there is no external force exerted on them.  Since Blood is a highly viscous medium,
the chance for the RBC to accelerate is negligible even in the presence of external force.  In the presence of external force (centrifugal forces), these cells undergo one-directional motion and they separate from the serum or plasma as the angular speed increases. The surface tension also affects the dynamics of the blood cells. This can be appreciated by substituting Eqs. (2) and (27) in Eq. (11). After some algebra  one gets
\begin{eqnarray}
V(t)&=&Nm\omega^2 r \left({1-e^{-{6r'\pi{(S-0.03699)\over  5.9125}t\over m}}\over 6r'\pi{(S-0.03699)\over  5.9125}}\right).
\end{eqnarray}
The position of the cell is then given by
$
x=\int_{0}^{t}V(t')dt'.
$
Exploiting Eq. (30), one can see that $V(t)$ or $x$  decreases as the surface tension steps up.

Moreover, the erythrocyte sedimentation rate depends on how the test tube is positioned. An inclination of the test tube by 3 degrees increases the ESR up to 30 percent \cite{aaaaa5}. Since blood is a highly viscous fluid,   its viscosity is strongly related to the surface tension as shown in Eq. (22).  On the other hand, the surface tension has a functional dependence on the degree of tilt.  When the test tube becomes tilted,  the surface tension of the blood decreases, and consequently its viscosity decreases.  As a result  a higher ESR is observed. 
	
\section {Summary and conclusion}

 In this paper,  via an exact analytical solution,  we study the factors 
that affect the efficacy of the centrifugation process. 
The effect of the centrifugation time on the efficacy of the centrifugation process is explored by studying the dynamics of the blood cells via the well-known Langevin equations or equivalently,  by solving Fokker-
Plank equations.  As blood cells are microscopic in size, their dynamics can be modeled as a Brownian particle walking in a viscous medium.  Since blood is a highly viscous medium, the chance for the blood cells to accelerate is negligible and the corresponding dynamics can be studied via the Langevin equation or Fokker Planck equation. Solving the Fokker Planck equation analytically, we explore how the dynamics of blood cells behave as a function of the model parameters. Because our study is performed by considering real physological  parameters, the results obtained in this work not only agree with the experimental observations but also help to understand most hematological experiments that are conducted in vitro.

In a medical laboratory, the standard test tube has a length of $L=150 mm$ and the cells separate from the plasma or serum at the result of the centrifugal force $f=Nm\omega^2 r $ where $N$  denotes the number of cells that form a cluster while $m$ designates the mass of $RBC$ or platelets.  Since the $WBCs$ are heavy, they move way before $RBCs$, and as an approximation, we disregard their dynamics.

The speed of the centrifuge
is one of the main factors concerning the efficacy of the centrifugation process. It is shown that as the angular speed increases, the centrifugal force steps up and as result, the particles are forced to separate
from the plasma or serum fast. Based on the size and fragility of the sample,  the centrifugation speed should be adjected. To increase the efficiency of the centrifugation process,  as the size of the particle decreases,  the centrifugation speed should step up.  For instance, our work depicts that the velocity for platelets is considerably lower than red blood cells due to the small size of platelets.  The  RBC  separates far more than platelets showing that to avoid the effect of clotting factors,  one has to increase the centrifugation time.  We also show that as the length of the test tube steps, the centrifugal force increases.
The dynamics of analyse during centrifugation is also affected by the room temperature. The generation of heat during centrifugation steps up the temperature within a centrifuge and as a result, not only the stability of the sample but also the mobility of the sample is affected.   Our result shows that as the centrifuge temperature steps up, the velocity of the cells as well as the average distance of the cell in the fluid decreases. This effect of temperature can be corrected by increasing the centrifugation time. However,  a longer centrifugation time might lead to a considerable amount of heat generation. This, in turn,  may cause the red blood cells to lyse. Prolonged centrifugation at high speed also causes structural damage to the cells and as a result, hemolysis occurs.  On the contrary, low-speed centrifugation leads to insufficient separation of plasma and serum from cellular blood components.  When the RBC forms rouleaux (as $N$ increases), the velocity and the position of the particle step up. This is because as $N$ increases, the centrifugal force increases.  This also implies since the RBC in serum forms aggregates due to clotting factors, it needs less centrifugation time than plasma.

The dynamics of the whole blood during capillary action is studied where in this case the blood flows upward in narrow spaces without the assistance of external forces. In this work, we first analyzed the correlation between
surface tension and viscous friction via data fitting. We show that the viscosity  steps up linearly as the surface tension increases.  The  mathematical relation between the height and viscous friction is derived. It is shown that 
the height the blood rises increases as the viscous friction increases. As the temperature of the room steps up, the height decreases.  
The dependence of the erythrocyte sedimentation rate (ESR) on model parameters is also studied.

Most medical laboratory examinations are affected by external factors such as temperature. To understand the factors that affect the outcome of these routine examinations,   it is vital to explore the dynamics of the whole blood, erythrocytes (RBCs), leukocytes (WBCs), and thrombocytes (platelets). Since our mathematical analysis is performed by using physiological parameters, all of the results depicted in this work can be reconfirmed experimentally. The simplified model presented in this work can also help to understand most hematological experiments that are conducted in vitro.

 {\it Acknowledgment.\textemdash} 
I would like to thank Mulu Zebene and Blyanesh Bezabih for the constant  encouragement.



 {\it Author contribution statements.\textemdash} Mesfin Taye conceived the research idea,  developed the theory, and performed the analytical computations. He also contributes to the writing of the manuscript.

\end{document}